\documentclass{appolb}
\usepackage{graphicx,amsmath,hyperref}
\usepackage{lineno}
\usepackage[utf8]{inputenc}

\begin{document}
\title{NA61/SHINE L\'evy HBT measurements in Be+Be collisions at 150A GeV/c
\thanks{Talk presented at XIII Workshop on Particle Correlations and Femtoscopy, 22-26 May 2018, Krak\'ow, Poland}%
}
\author{Barnab\'as P\'orfy for the NA61/SHINE collaboration
\address{E\"otv\"os Lor\'and University, H-1117 Budapest, P\'azm\'any P. s. 1/A, Hungary}
\\
}
\maketitle
\begin{abstract}
The program of NA61/SHINE allows for the investigation of the phase diagram of strongly interacting matter. The nature of the quark-hadron transition can be studied through analyzing the space-time structure of the hadron emission source via the measurement of Bose-Einstein momentum correlations. These can be described  by correlation functions based on L\'evy-distributed sources. This report presents the performance plots of Bose-Einstein correlation analysis in Be+Be collisions at 150$A$ GeV$/c$ beam momentum. The transverse mass dependence of the L\'evy  source parameters and their possible interpretations are discussed.	
\end{abstract}
  
\section{Introduction}
NA61/SHINE is a fixed target experiment located in the CERN SPS North Area. It has 4 large time projection chambers (TPCs), two of which are placed inside strong vertex magnets. The TPCs are covering the full forward hemisphere and with precise tracking down to a $p_T$ of 0 GeV$/c$. NA61/SHINE also has a Projectile Spectator Detector (PSD), a modular calorimeter located on the beam axis, which measures the forward energy $E_F$. This allows to determine the centrality of the collisions. 

In NA61/SHINE various nuclei are collided (proton-proton, proton-nucleus and nucleus-nucleus) at multiple energies (13-150$A$ GeV$/c$). The phase diagram of strongly interacting matter is being investigated in numerous experiments at many accelerators and colliders. If there is a critical point, NA61/SHINE has a good chance to observe its signatures. The QCD phase diagram can be studied by analyzing data at different baryon-chemical potential ($\mu_B$) and temperature ($T$). One important method is to analyze the data by measurements of Bose-Einstein (HBT) correlations. In this paper performance plots will be shown of such an analysis in Be+Be collisions at 150$A$ GeV$/c$.

\section{Search for CEP with L\'evy type of HBT analysis}
A critical point always has to be characterized by its critical exponents. One of these exponents is $\eta$ which is related to spatial correlations. At the critical point fluctuations appear at all scales and the spatial correlation function becomes a power-law $\sim r^{-(d-2+\eta)}$ (where $d$ represents the number of dimensions). The exponent $\eta$ of this power-law can be conjectured based on the universality class, which for QCD is that of the the 3D Ising model~\cite{Halasz:1998qr,Stephanov:1998dy}. With this assumption one can predict $\eta$ at the CEP of QCD. From the 3D Ising model, this value is $\eta = 0.03631$~\cite{El-Showk2014}, while the 3D Ising model with random external field yields the exponent $\eta$ is at $0.5 \pm 0.05$~\cite{Rieger:PhysRevB.52.6659}.

These spatial momentum correlations can be investigated by measuring Bose-Einstein correlations. R. Hanbury Brown and R. Q. Twiss (HBT method) observed intensity correlations as a function of detector distance while observing Sirius with optical telescopes~\cite{HanburyBrown:1956bqd}. This method enabled them to measure the size of point like sources. Goldhaber and collaborators observed momentum correlations by adapting this principle to high energy particle physics~\cite{Goldhaber:1959mj}. This led to the observation that the momentum correlation function $C(q)$
is directly related to the normalized source distribution $S(r)$ via
\begin{equation}
C(q) \cong 1 + | \tilde{S} |^2,
\end{equation}
where $\tilde{S}(q)$ is the Fourier transform of $S(r)$.
Nowadays one uses this technique to extract the femtometer scale structure of hadron production in heavy ion collisions. The shape of this source is usually assumed to be Gaussian,
however alternative approaches might be more realistic. The expanding medium increases the mean free path which may lead to anomalous diffusion and L\'evy distributed sources~\cite{METZLER20001,Csanad:2007fr}. Hence it is interesting to check if the source indeed looks like a L\'evy distribution. The L\'evy distribution is defined as:
\begin{equation}
\mathcal{L}(\alpha,R,r)=\frac{1}{(2\pi)^3} \int d^3q e^{iqr} e^{-\frac{1}{2}|qR|^{\alpha}}.
\end{equation}
The L\'evy distribution is analytically calculable for $\alpha=1$ and 2, and it yields a Cauchy or a Gaussian distribution, respectively.
One important difference between an $\alpha<2$ L\'evy and a Gauss distribution is the power-law tail $\sim r^{-(d-2+\alpha)}$ (where $d$ is the number of spatial dimensions). Hence L\'evy distributions lead to spatial correlation functions with the same power-law tails (which is why the L\'evy distributions are called stable distributions). The spatial correlation at the critical point is $\sim r^{-(d-2+\eta)}$. This immediately suggests that the L\'evy exponent $\alpha$ is identical to the spatial correlation exponent $\eta$~\cite{Csorgo:2003uv}. For the value conjectured for the CEP, $\alpha$ has to be at or below 0.5. In general, anomalous diffusion alone leads to $\alpha<2$, but in the vicinity of the CEP, very low $\alpha$ values around 0.5 may be expected. This can be measured by investigating Bose-Einstein correlation functions. For these, the result (based on L\'evy distributions) is as follows:
\begin{equation}
C(q) = 1 + \lambda \cdot e^{-(qR)^\alpha}.\label{e:levyC0}
\end{equation}
One may observe that if the shape parameter $\alpha$ is 2, then $C(q)$ is Gaussian, if $\alpha = 1$ then $C(q)$ is exponential. Such correlation function shapes were observed by PHENIX in Ref.~\cite{Adare:2017vig}.

\section{Details of the analysis}
This paper reports measurements performed in Be+Be collisions at $150A$ GeV$/c$. Performance results are shown for a selection of events with centrality higher than 20\%, where centrality was defined using the Projectile Spectator Detector. Due to non-uniform efficiency in this range, this event selection may be prone to trigger bias. Hence results of this analysis should be regarded as performance results. Event and track quality cuts were applied on this sample and negatively charged tracks were selected. Even though particles were not identified, the applied cuts already result in a relatively clean pion sample, since particle contamination from kaons is relatively low: kaon multiplicities in proton-proton collisions at the same energy are less than 2\% of the pion multiplicity~\cite{Aduszkiewicz:2017sei}. Analysis of positively charged particles and identified hadrons is the subject of a subsequent analysis. Pair distributions were measured as a function of the 1D momentum difference in the longitudinally comoving system:
\begin{align}
|q_{LCMS}| &= \sqrt{q^2_x + q^2_y + q^2_{z,LCMS}},\textnormal{ where}\\
q^2_{z,LCMS} &= 4\cdot \frac{(p_{z,1} \cdot E_2 - p_{z,2} \cdot E_1)^2}{(E_1 + E_2)^2 - (p_{z,1} + p_{z,2})^2}.
\end{align}
This was found to be the correct 1D variable in this system, based on the analysis of multidimensional correlation functions. Pairs were grouped according to the average transverse momentum ($K_T$) of the pair, defining 4 $K_T$ bins ranging from 0-600 MeV$/c$. In each bin the measured correlation function was measured as the ratio of the distributions of real and mixed pairs, similarly to e.g.~\cite{Adare:2017vig}. 

The repulsion of the members of pairs of like-sign charge particles has to be taken into account in the analysis. The standard method for this is the application of a Coulomb-correction. It is a complicated numerical procedure to calculate the Coulomb correction for a L\'evy source, and to fit the correlation function with the result of such a calculation. It can be observed however, that the Coulomb-correction does not depend strongly on the value of the shape parameter $\alpha$ in Eq.~(\ref{e:levyC0}). It is also important to note, that the Be+Be system is quite small, so the Coulomb effect is not as important as for example in Pb+Pb. Hence it is well justified to use the formula published by CMS~\cite{Sirunyan:2017ies}, which is an approximation for $\alpha = 1$:
\begin{align}
K_{\textrm{Coulomb}}(q) &= \textrm{Gamow}(q)\cdot \left(1+\frac{\pi\eta q\frac{R}{\hbar c}}{1.26+q \frac{R}{\hbar c}}\right),\textnormal{ where}\label{e:coulcorr}\\
\textrm{Gamow}(q) &= \frac{2\pi\eta(q)}{e^{2\pi\eta(q)-1}} \;\textnormal{ and }\;
\eta(q) = \alpha_{\textrm{QED}}\cdot\frac{\pi}{q}.
\end{align}

\section{Results and conclusions}
The measured correlation functions were fitted with the result of Eq.~\eqref{e:levyC0}, taking into account also the Coulomb correction of Eq.~\eqref{e:coulcorr}. In the fit function, there are three important fit parameters: $R, \lambda$ and $\alpha$. They are measured in 4 $K_T$ bins, as mentioned above, and one may investigate their dependence on the transverse mass $m_T$ value corresponding to the given $K_T$ bin.

The first of the fit parameters, the L\'evy scale $R$ determines the length of homogeneity. In a simple hydrodynamical model one obtains an $1/R^2\propto m_T$ type of dependence, creating a decreasing trend with $m_T$ which can be caused by the transverse flow. It is an important question if this can be observed in Be+Be collisions as well, as such an effect was also found in RHIC p+p collisions~\cite{Aggarwal:2010aa}.
The correlation strength parameter $\lambda$ can be interpreted in the core-halo picture, where the core consists of pions created close to the center, i.e. primordial pions or the decay products of short lived resonances, while the halo contains the decay pions from long lived resonances. In this picture, $\lambda$ can be expressed as
\begin{align}
\lambda(m_T) = \left(\frac{N_{core}}{N_{core}+N_{halo}}\right)^2.
\end{align}
This parameter was found to show a low-$m_T$ decrease in Au+Au collisions at RHIC~\cite{Abelev:2009tp,Adare:2017vig}, but no such tendency was found in Pb+Pb reactions at the SPS~\cite{Beker:1994qv,Alt:2007uj}. Hence, it is interesting to investigate this phenomenon in other collision systems.
Finally the L\'evy exponent $\alpha$ may show a signal for anomalous diffusion, if $\alpha<2$ is found, similarly to Ref.~\cite{Adare:2017vig}, where $\alpha\approx 1.2$ was found, approximately independently of $m_T$. In the context of the search for the critical endpoint, it is even more important to investigate if $\alpha$ is in the vicinity of the conjectured value of 0.5.

\begin{figure}
\vspace{-7pt}
\centerline{%
\includegraphics[width=0.7\linewidth]{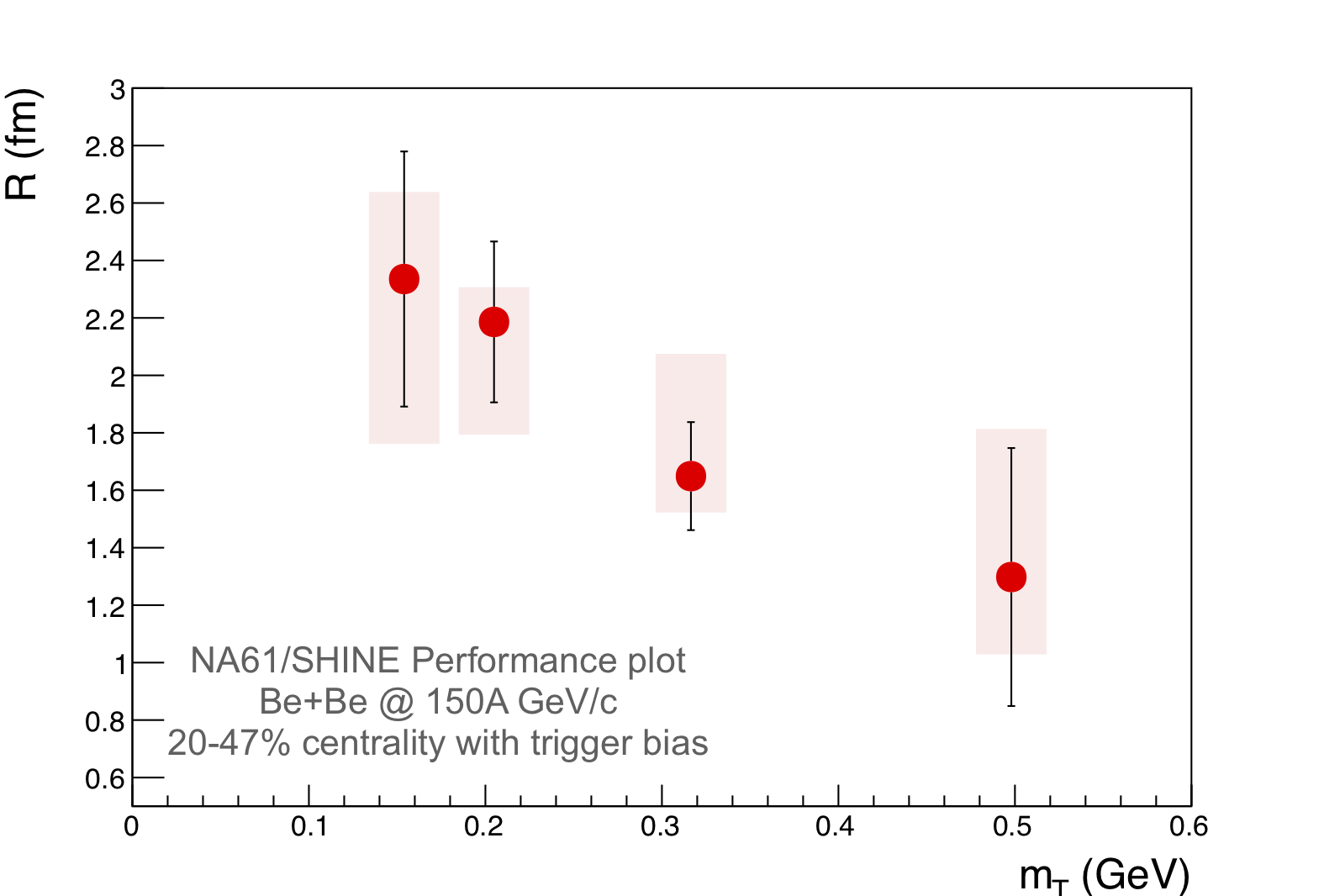}}
\vspace{-7pt}
\caption{Fit parameter $R$ versus average $m_T$ of the pair with systematic and statistical uncertainties, illustrated with box and bars, respectively.}
\label{f:R}
\end{figure}

\begin{figure}
\vspace{-7pt}
\centerline{%
\includegraphics[width=0.7\linewidth]{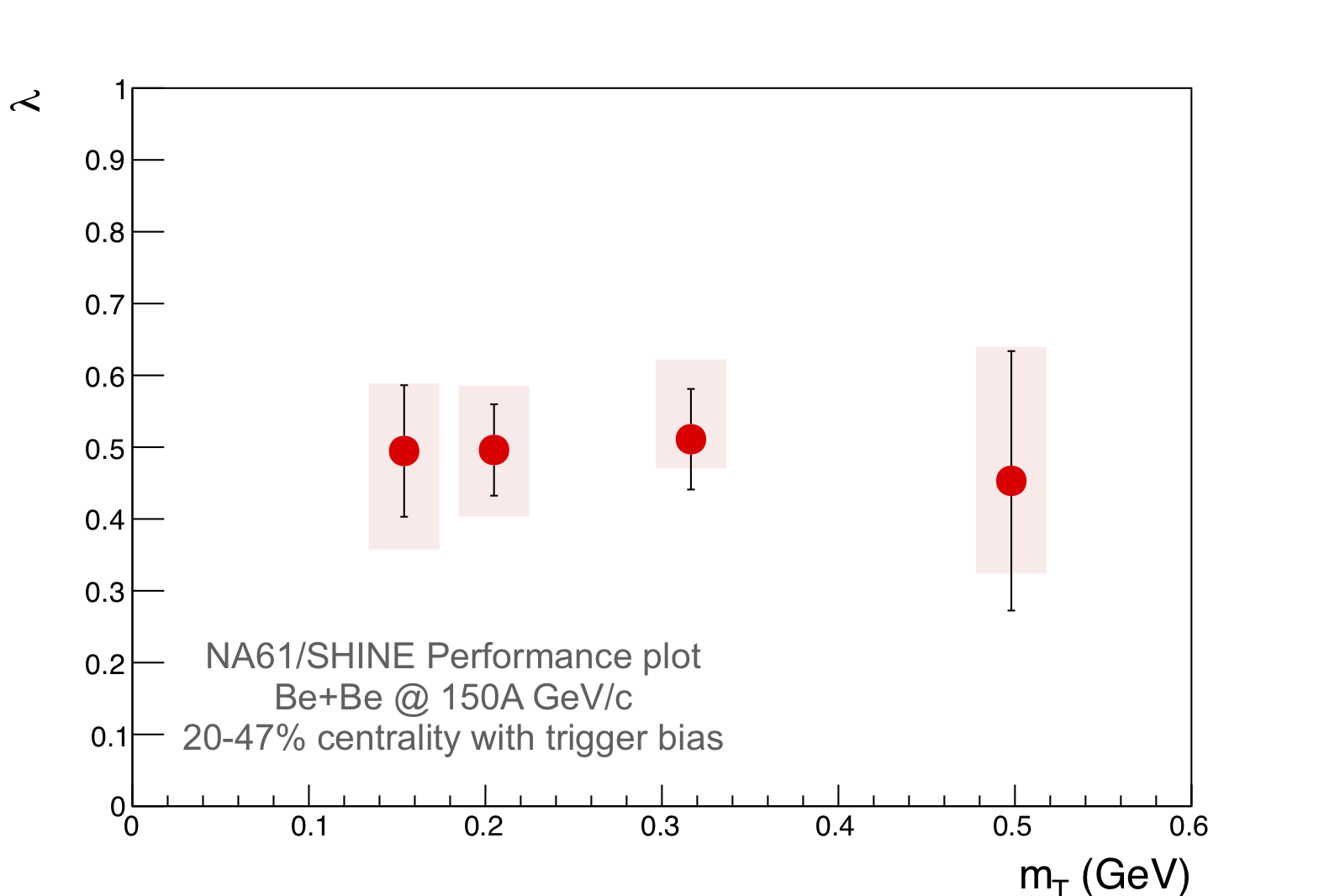}}
\vspace{-7pt}
\caption{Fit parameter $\lambda$ versus average $m_T$ of the pair with systematic and statistical uncertainties, illustrated with box and bars, respectively.}
\label{f:L}
\end{figure}

\begin{figure}
\vspace{-7pt}
\centerline{%
\includegraphics[width=0.7\linewidth]{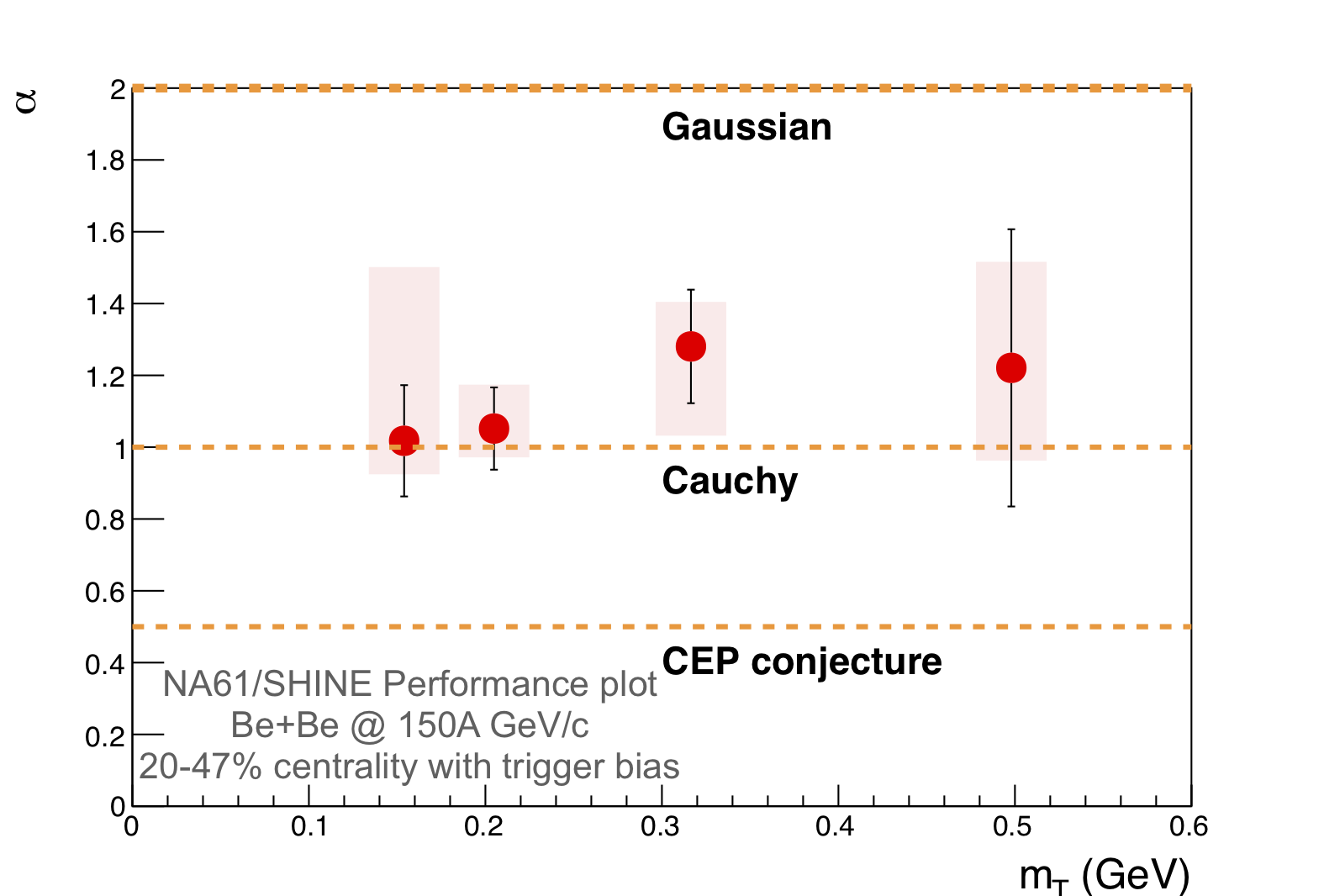}}
\vspace{-7pt}
\caption{Fit parameter $\alpha$ versus average $m_T$ of the pair with systematic and statistical uncertainties, illustrated with box and bars, respectively.}
\label{f:A}
\end{figure}

Figs.~\ref{f:R}, \ref{f:L} and \ref{f:A} show performance plots on the $m_T$ dependence of the above mentioned parameters in mid-central Be+Be collisions at 150$A$ GeV$/c$ beam momentum. These results clearly demonstrate the capabilities of NA61/SHINE to measure BE correlations even in this low multiplicity collision system. It will be an important and interesting step to measure these observables in other collision systems, as this may shed light on a number of interesting phenomena in collision systems of different sizes, such as radial flow and the vicinity of the critical point.

\section*{Acknowledgments}
The author expresses thanks to the NA61/SHINE collaboration, as well as NKFIH grants FK123842 and FK123959.


\begin{thebibliography}{10}

\bibitem{Halasz:1998qr}
A. Halasz {\it et~al.},
\newblock Phys. Rev. D58 (1998) 096007.

\bibitem{Stephanov:1998dy}
M. Stephanov, K. Rajagopal and E. Shuryak,
\newblock Phys. Rev. Lett. 81 (1998) 4816.

\bibitem{El-Showk2014}
S. El-Showk {\it et~al.},
\newblock Journal of Statistical Physics 157 (2014) 869.

\bibitem{Rieger:PhysRevB.52.6659}
H. Rieger,
\newblock Phys. Rev. B 52 (1995) 6659.

\bibitem{HanburyBrown:1956bqd}
R. Hanbury~Brown and R.Q. Twiss,
\newblock Nature 178 (1956) 1046.

\bibitem{Goldhaber:1959mj}
G. Goldhaber et~al.,
\newblock Phys. Rev. Lett. 3 (1959) 181.

\bibitem{METZLER20001}
R. Metzler and J. Klafter,
\newblock Physics Reports 339 (2000) 1 .

\bibitem{Csanad:2007fr}
M. Csan\'ad, T. Cs\"org\H{o} and M. Nagy,
\newblock Braz. J. Phys. 37 (2007) 1002.

\bibitem{Csorgo:2003uv}
T. Cs\"org\H{o}, S. Hegyi and W.A. Zajc,
\newblock Eur. Phys. J. C36 (2004) 67.

\bibitem{Adare:2017vig}
A. Adare et~al. [PHENIX Coll.],
\newblock Phys. Rev. C97 (2018) 064911.

\bibitem{Aduszkiewicz:2017sei}
A. Aduszkiewicz {\it et~al.} [NA61/SHINE Coll.],
\newblock Eur. Phys. J. C77 (2017) 671.

\bibitem{Sirunyan:2017ies}
A.M. Sirunyan {\it et~al.} [CMS Coll.],
\newblock Phys. Rev. C97 (2018) 064912.

\bibitem{Aggarwal:2010aa}
M.M. Aggarwal {\it et~al.} [STAR Coll.],
\newblock Phys. Rev. C83 (2011) 064905.

\bibitem{Abelev:2009tp}
B.I. Abelev {\it et~al.} [STAR Coll.],
\newblock Phys. Rev. C80 (2009) 024905.

\bibitem{Beker:1994qv}
H. Beker {\it et~al.} [NA44 Coll.],
\newblock Phys. Rev. Lett. 74 (1995) 3340.

\bibitem{Alt:2007uj}
C. Alt {\it et~al.} [NA49 Coll.],
\newblock Phys. Rev. C77 (2008) 064908.

\end{thebibliography}

\end{document}